\documentclass[sigconf]{acmart}

\usepackage{tikz}
\usepackage{pgf-umlcd}
\usepackage[edges]{forest}
\usetikzlibrary{arrows.meta,shadows.blur}
\colorlet{myyellow}{yellow!50}
\usepackage{float}
\usepackage[ruled,vlined]{algorithm2e}
\usepackage{dblfloatfix}

\usetikzlibrary{shapes,arrows}
\usepackage{amsmath,bm,times}
\usepackage{pgfplots}
\usepackage{pgfplotstable}
\usepackage{subcaption}
\usepackage{cleveref}
\usepackage[export]{adjustbox}
\usepackage{bera}
\usepackage{listings}
\usepackage{xcolor}

\colorlet{punct}{red!60!black}
\definecolor{background}{HTML}{EEEEEE}
\definecolor{delim}{RGB}{20,105,176}
\colorlet{numb}{magenta!60!black}

\usepackage{pythonhighlight}
\definecolor{comment-text-color}{rgb}{0,0.8,0.6}
\lstdefinelanguage{json}{
    basicstyle=\normalfont\ttfamily\footnotesize,
    numbers=left,
    numberstyle=\scriptsize,
    stepnumber=1,
    numbersep=8pt,
    showstringspaces=false,
    breaklines=true,
    frame=lines,
    backgroundcolor=\color{background},
    literate=
     *{0}{{{\color{numb}0}}}{1}
      {1}{{{\color{numb}1}}}{1}
      {2}{{{\color{numb}2}}}{1}
      {3}{{{\color{numb}3}}}{1}
      {4}{{{\color{numb}4}}}{1}
      {5}{{{\color{numb}5}}}{1}
      {6}{{{\color{numb}6}}}{1}
      {7}{{{\color{numb}7}}}{1}
      {8}{{{\color{numb}8}}}{1}
      {9}{{{\color{numb}9}}}{1}
      {:}{{{\color{punct}{:}}}}{1}
      {,}{{{\color{punct}{,}}}}{1}
      {\{}{{{\color{delim}{\{}}}}{1}
      {\}}{{{\color{delim}{\}}}}}{1}
      {[}{{{\color{delim}{[}}}}{1}
      {]}{{{\color{delim}{]}}}}{1},
}

\relax

\AtBeginDocument{%
  \providecommand\BibTeX{{%
    \normalfont B\kern-0.5em{\scshape i\kern-0.25em b}\kern-0.8em\TeX}}}

\acmConference[IEEE Transactions on Computers: Special Issue on Quantum Computing]
\begin{document}

\title{Enabling Pulse-level Programming, Compilation, and Execution in XACC}

\author{Thien Nguyen}
\affiliation{%
  \institution{Computer Science and Mathematics \\ Oak Ridge National Laboratory}
}
\email{nguyentm@ornl.gov}

\author{Alexander McCaskey}
\affiliation{%
  \institution{Computer Science and Mathematics \\ Oak Ridge National Laboratory}
%   \country{Iceland}
}
\email{mccaskeyaj@ornl.gov}

\begin{abstract}
Noisy gate-model quantum processing units (QPUs) are currently available from vendors 
over the cloud, and digital quantum programming approaches exist to run low-depth circuits on physical hardware. These 
digital representations are ultimately lowered to pulse-level instructions by vendor quantum control 
systems to affect unitary evolution representative of the submitted digital circuit. 
Vendors are beginning to open this pulse-level control system to the public via specified interfaces. 
Robust programming methodologies, software frameworks, and backend simulation technologies for 
this analog model of quantum computation will prove critical to advancing pulse-level control research and development. 
Prototypical use cases for this include error mitigation, optimal pulse control, and physics-inspired pulse construction. 
Here we present an extension to the XACC quantum-classical software framework that enables 
pulse-level programming for superconducting, gate-model quantum computers, and a novel, general, 
and extensible pulse-level simulation backend for XACC that scales on classical compute clusters via MPI. 
Our work enables custom backend Hamiltonian definitions and gate-level compilation to available 
pulses with a focus on performance and scalability. We end with a demonstration of this capability, and 
show how to use XACC for pertinent pulse-level programming tasks.

% As low-level pulsing control of quantum hardware which becomes available to developers, 
% the need arises for integrated and holistic software 
% frameworks that can cope with this extended quantum programming model. Here we present
% an extension to the XACC heterogeneous quantum-classical computing framework to enable
% pulse-level quantum programming and simulation in a multi-layered and hardware-agnostic
% approach. This enables us to represent quantum hardware back-end at the Hamiltonian level and 
% to compile gate-level quantum circuits into pulses which can then be simulated dynamically
% or executed on pulse-capable hardware back-ends. We implemented a simulator back-end 
% with a strong focus on performance and scalability. We provide concrete examples of how
% to register pulse-capable systems to the framework and of various pulse utilities 
% that it provides through this extension.  
\end{abstract}

\maketitle

\section{Introduction}\label{Introduction}

% FIXME Perhaps work this back in - In this manuscript, we will present our recent work on the XACC framework to support pulse-level quantum programming.
% Specifically, we enhanced our IR model to represent analog instructions, implemented a vendor-agnostic pulse instruction
% scheduler, and developed an in-house HPC (high-performance computing) oriented pulse-level simulator back-end.
% The well-designed IR infrastructure (e.g. compiler front-ends, composite IR's, IR transformations) and our microservices-based architecture 
% have enabled seamless integration of these new functionalities. In the accompanying demonstration examples, we will show
% how an existing gate-based quantum program being transformed into pulses and being simulated dynamically.
% More importantly, advanced users can now, via XACC, implement and test fully customizable gates/circuits to pulses mappings
% before deployment. 

The emergence of noisy intermediate scale quantum (NISQ) \cite{preskill2018quantum} processing units and their accessibility to programmers via cloud services in recent years have led to major advances in software frameworks and programming models for quantum-classical computing. For gate-model quantum computing the majority of the programming framework approaches have focused on digital assembly submission to dedicated quantum job servers. This has led to a number of demonstrations of small-scale experiments in machine learning, quantum chemistry, nuclear physics, and quantum field theory \cite{hamilton,Dumitrescu2018,McCaskey2019,PhysRevA.98.032331,OMalley2016,Kandala2017}. However, in the NISQ-era, much stands to be gained from thinking about programming quantum hardware closer to the physical model of the system, specifically, programming at an analog, or pulse-instruction, level. Early results in this regard have demonstrated the utility of pulse-level control in areas such as error mitigation and optimal quantum control \cite{Kandala2019,chong_control_grape}. Moreover, vendors are starting to propose and implement specifications for pulse-level job submission, opening up this novel, analog programming model to the community at large. In the near-term, programming models and software framework implementations will prove critical in enabling the benefits of low-level analog quantum control and programming. Moreover, robust simulation technologies are required to promote pulse-level research productivity due to the scarce and remote characteristic of available pulse-level physical backends. 

Programming mechanisms for pulse-level quantum computation is very much at an early research and development phase. IBM has recently put forward the OpenPulse specification, with a corresponding early implementation in the Qiskit-Terra framework \cite{mckay2018qiskit}. This implementation provides a high-level Pythonic approach for pulse-level program composition and specifically targets IBM physical backends. Pulse-level simulation tools exist but are available in an ad-hoc manner - most users leverage frameworks such as QuTip for quick prototyping and scripting \cite{qutip}. These approaches lack integration with physical hardware backends and internal circuit representations for the typical quantum programming, compilation, and execution workflow. There is a need for a system-level programming, compilation, and execution workflow that enables scalable simulation on available distributed-memory compute architectures. 
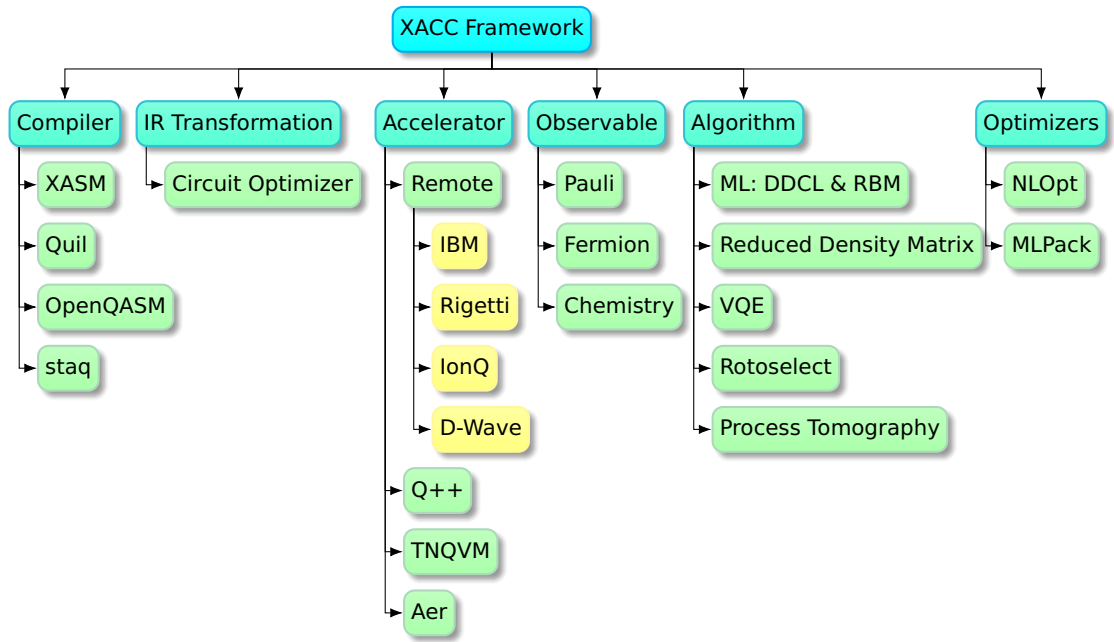
\begin{figure*}[t!] 
  \begin{forest}
    colour me/.style={top color=#1!75, bottom color=#1, draw=#1, thick, blur shadow, rounded corners},
    for tree={
      edge=-Latex,
      font=\sffamily,
    },
    where level=1{
      for tree={
        folder,
        grow'=0,
      },
      edge path'={(!u.parent anchor) -- ++(0,-15pt) -| (.child anchor)},
    }{},
    before typesetting nodes={
      for tree={
        content/.wrap value={\strut #1},
      },
      if={isodd(n_children("!r"))}{
        for nodewalk/.wrap pgfmath arg={{fake=r,n=#1}{calign with current edge}}{int((n_children("!r")+1)/2)},
      }{},
      tempcounta/.max={level}{tree},
      for tree={
        colour me/.wrap pgfmath arg={cyan!#1!myyellow}{100*((tempcounta)-level())/(tempcounta)}
      }
    }
    [XACC Framework
      [Compiler
        [XASM]
        [Quil]
        [OpenQASM]
        [staq]
      ]  
      [IR Transformation
        [Circuit Optimizer]
      ]
      [Accelerator
        [Remote
          [IBM]
          [Rigetti]
          [IonQ]
          [D-Wave]
        ]
        [Q++]
        [TNQVM]
        [Aer]
      ]
      [Observable
        [Pauli]
        [Fermion]
        [Chemistry]     
      ]
      [Algorithm
        [ML: DDCL \& RBM]
        [Reduced Density Matrix]
        [VQE]
        [Rotoselect] 
        [Process Tomography]   
      ]
      [Optimizers
        [NLOpt]
        [MLPack]
      ]
    ]
  \end{forest}
  \caption{The XACC framework provides a service-oriented architecture. Services are interfaces that expose core functionality for the typical quantum-classical programming, compilation, and execution workflow. Here we demonstrate the hierarchy of these services with concrete service implementations.}
  \label{fig:xacc_components}
  \end{figure*}
  
Here we detail our solution to the current lack of programming tools and simulation technologies for pulse-level programming of gate-model quantum architectures. We leverage the system-level XACC heterogeneous quantum-classical software framework to provide a programming mechanism and associated compilation tooling for pulse-level quantum control systems. Initially we target the OpenPulse specification, but believe that the hardware-agnostic characteristic of XACC will enable us to extend this model for future quantum control specifications (such as Rigetti Quilt). We put forward a state-of-the-art pulse-level simulation backend for XACC that builds off the QuaC (Quantum in C) open quantum system dynamics solver. QuaC builds off of the PETSc matrix algebra library, and therefore our work immediately scales on distributed memory architectures via the Message Passing Interface (MPI). We detail programming, compilation, and execution layers of this overall pulse-level programming model, and provide detailed demonstrations exhibiting the use of this API. Our goal with this work is to provide a unique pulse-level programming capability with a robust simulation backend to enable research productivity for tasks in error mitigation, physics-inspired pulse learning, and quantum optimal control.

% \paragraph{Outline}
The paper is organized as follows. In Section~\ref{XACC_Framework}, we review some key features and components
of the XACC framework. Next, in~\ref{OpenPulse}, we briefly summarize the OpenPulse specification which we adopt
in this work. Section~\ref{Software-Architecture} describes the overall software architecture of our pulse-level extension
and how it fits in with the framework. Usage demonstrations and conclusions are given in Section~\ref{Demonstration}
and Section~\ref{Discussion}, respectively.

\section{XACC Framework} \label{XACC_Framework}

Recent progress in the development of programmable quantum computing architectures has resulted in a number of programming models and associated software stack implementations enabling digital gate-level programming of these nascent architectures. Most efforts have been put forward by hardware vendors and typically only target the backend architecture the vendor provides. Most of these efforts are provided via high-level Pythonic circuit construction data structures, with a few specifically extending existing classical languages for heterogeneous quantum-classical computing. This landscape has necessitated the development of unified, system-level architectures for robust integration efforts across quantum architectures, languages, and compiler techniques \cite{McCaskeyICRC2018}. 

\begin{table*}[!b]
\begin{center}
\begin{tabular}{ | p{0.2\textwidth} | p{0.4\textwidth}| p{0.35\textwidth} | } 
\hline
Type  & Definition & Examples \\ 
\hline
Operators & - Pauli operators (\texttt{I}, \texttt{X},  \texttt{Y},  \texttt{Z})\newline
		     - Ladder operators (\texttt{SM}, \texttt{SP})\newline 
		     -	 Number operators  (\texttt{O}, \texttt{N}) 	&  \texttt{2*pi*v0*Z0} \newline
													    \texttt{2*pi*v0*SP0*SM0}\newline
													    \texttt{0.5*delta0*(1-O0)*O0}  \\
\hline
Drive and Control Channel & Special suffix (\texttt{||D} or \texttt{||U}) 
						to denote the Hamiltonian term is modulated 
						by either a drive or control channel 	& \texttt{g*X0||D0} \\ 
\hline
Summation & A running sum of operator expression indexed by an integer 	&\texttt{\_SUM[i,0,9,-0.5*h\{i\}*Z\{i\}]}
																	   \texttt{\_SUM[i,0,9,g\{i\}*X\{i\}||D\{i\}]} \\ 
\hline
Variables & A map from symbolic variables to values	& \texttt{``vars'':\{``v'': 5.0, ``g'': 0.1\}}\\ 
\hline
Qubit dimensions & A map from qubit indices to dimensions & \texttt{``qub'':\{``0'': 2, ``1'': 2\}}\\
\hline
\end{tabular}
\end{center}
\caption{JSON schema of OpenPulse backend Hamiltonian specification}
\label{tab:openpulse_table}
\end{table*}
Recently, the XACC system-level software framework has been developed to address these requirements. XACC provides an extensible, holistic framework for quantum-classical computing in C++ in a way that is agnostic to the backend quantum hardware type as well as front-end language transpilation and compilation \cite{mccaskey2020xacc}. 
At its core, XACC provides a unified and simplified quantum programming model
that can cross hardware boundaries while not compromising efficiency in terms of quantum hardware 
utilization or developers productivity. Moreover, the system-level language approach ensures performance of classical processing and support code which
is essential in the NISQ-era where hybrid algorithms are one of most viable approaches that can 
provide substantial quantum advantage.

As a comprehensive framework, XACC (see Fig.~\ref{fig:xacc_components}) 
adheres to the newly proposed QCOR language extension specification \cite{mintz2019qcor}, which provides guidance on the programming and memory model,
support data structures, and the execution model. XACC provides a reference implementation that is open-source, fully-functional, actively being maintained and developed \cite{eclipse-xacc}.
The base implementation supports all major quantum assembly dialects, e.g. Open QASM, Quil, and Staq \cite{openqasm,quil,staq}, as well as its own 
XASM dialect. XACC has back-end connections, e.g. via RESTful APIs, to hardware platforms provided by IBM, Rigetti, IonQ, and Dwave, thereby enabling a \emph{write-once-deploy/test-everywhere} capability to programmers and researchers. Another key focus area is the integration of quantum simulation software to use as drop-in replacements for real hardware back-end during development.
XACC support simulation backends from the IBM Terra/AER package \cite{Qiskit}, the Quantum\texttt{++} library \cite{qpp}, the ITensor library \cite{tnqvm-plos-one}, and the high-performance ExaTN tensor library \cite{exatn}.
This level of modularity and extensibility, and the overall strong focus on performance, significantly enhances developer productivity.

XACC provides an extensive set of libraries that are constantly updated. This includes algorithm implementations, from simple static algorithms,
e.g. quantum Fourier transform, reversible classical computing, to novel quantum variational ansatz circuits, such as UCCSD, ASWAP, 
or quantum circuit structure learning. Observables, such as Pauli or spin operators, are a native construct within the framework, and enable XACC to
automatically inject gates and measurement operations in order to compute pertinent expectation values. Moreover, XACC supports 
common hybrid algorithms like the variational quantum eigensolver and data-driven circuit learning routines, as well as characterization 
tools such as quantum process tomography. Again, these libraries are meant to provide developers
with drop-in, backend-compatible, and fully-validated solutions without needing them to implement or maintain the code. 

Lastly, XACC provides user-friendly Python language bindings via Pybind11 \cite{pybind11}. This is not a specification-driven feature but is equally important since
C++, despite being highly performant, requires substantial familiarity with professional software development procedures,
e.g., building and linking, that may represent a barrier to domain computational scientists. 
Thanks to the language bindings, users can use XACC services from their Python shell and, more importantly, can provide
implementations to framework constructs, e.g. Accelerators, Compilers, Algorithms, etc., natively from Python while adhering to
the common interfaces that have been exposed via the language bindings.

% \begin{itemize}

% \end{itemize}

\section{OpenPulse Specification} \label{OpenPulse}

The OpenPulse specification~\cite{mckay2018qiskit} was proposed by IBM as an open and implementation agnostic standard for pulse-level control. The specification defines JSON schemas for system Hamiltonians, drive and control channel
configurations, pulses, and pulse-sequence definitions. Since this is the first pulse-level standard put forward in the industry, our XACC implementation adheres to the OpenPulse specification.

\subsection{Hamiltonian} \label{hamiltonian}
The specification defines a machine-readable encoding scheme for the system Hamiltonian as summarized in Table \ref{tab:openpulse_table}.
Specifically, the Hamiltonian field (\lstinline{h_str} in the OpenPulse JSON object) is a list of strings describing various terms in the Hamiltonian.
These terms can be time-independent (control-free evolution) or time-dependent (driven by an input channel). 
The latter case is denoted by the characters \lstinline{||} followed by a channel name 
(either a drive channel, \lstinline{Di}, or a control channel, \lstinline{Ui}).

Also, there is a set of pre-defined operators, such as Pauli operators (\lstinline{X}, \lstinline{Y}, or \lstinline{Z}) 
or ladder operators (e.g. \lstinline{SM} denotes the spin $\sigma_{-}$ or the photon destroy ($a$) operator), that can be used to construct those Hamiltonian terms. For brevity, the standard also allows running summation terms with integer indices to group terms that have the same structure. 

All variables in the Hamiltonian expressions must have corresponding assignments in the \lstinline{vars} map to resolve the Hamiltonian dynamics. Additionally, qubit subsystems can be modeled as higher-dimensional systems (e.g. qutrits or qudits) by specifying their dimensions in the \lstinline{qub} field.

\subsection{Pulses and Commands}

Pulses (waveforms of drive signals) are defined as sequences of sample points (complex numbers). They are
specified as a JSON object with a \lstinline{name} and a \lstinline{samples} field (see Fig.~\ref{fig:pulse_cmd_def}).
% \begin{figure}[h!] 
% \begin{lstlisting}[language=json]
% {
% 	"name": "XACC_pulse",
% 	"samples": [[-0.18, -0.77], [-0.44, -0.18], ... ]
% }
% \end{lstlisting}
% \caption{JSON description of a pulse}
% \label{fig:pulse_json}
% \end{figure}
A standard OpenPulse back-end configuration often contains a calibrated pulse library and a map from quantum gates to sequences of pulses. Each mapping entry is called a command definition, or \lstinline{cmd-def}. A sample \lstinline{cmd-def} JSON entry is also shown in Fig.~\ref{fig:pulse_cmd_def}. 
\begin{figure}[!t] 
\begin{lstlisting}[language=json]
pulse : {
	"name": "XACC_pulse",
	"samples": [[-0.18, -0.77], [-0.44, -0.18], ... ]
}
cmd-def: {
	"name": "cx",
	"qubits": [0, 1],
	"sequence": [{
		"ch": "d0",
		"name": "fc",
		"phase": 1.5707963267948966,
		"t0": 0
	}, {
		"ch": "d0",
		"name": "Pulse1",
		"t0": 0
	}, ...
	]
}
\end{lstlisting}
\caption{JSON description of a simple pulse and a quantum gate consisting of multiple pulses in a sequence (cmd-def).}
\label{fig:pulse_cmd_def}
\end{figure}

As we can see, at the pulse level, a digital quantum gate is decomposed into a sequence of pulses (referred to by name)
and frame changes (changing the local oscillator's phase) on specific channels at specific time.
The pulse sequence for each gate is defined independently referencing the gate start time as time zero. 
A back-end pulse library can implement an arbitrary set of universal gates and hence it is up to
the high-level gate compiler to perform any necessary transformations and mappings.

\section{Software Architecture} \label{Software-Architecture}

The extensible, layered and future-proof interfaces that the XACC framework implements provide a foundation upon which our pulse-level extension is built. 
We supplement the front-end intermediate representation with a pulse instruction extension and implement a middle-end pipeline to enable gate-to-pulse translations. The processed pulse-level program can then be executed on the QuaC simulator/accelerator backend which performs a fully dynamical simulation of the quantum system driven by those pulse instructions. Of course, with backend support for quantum hardware from IBM, XACC can take this pulse-level IR and execute it on physical architectures as well.

\subsection{Pulse IR}

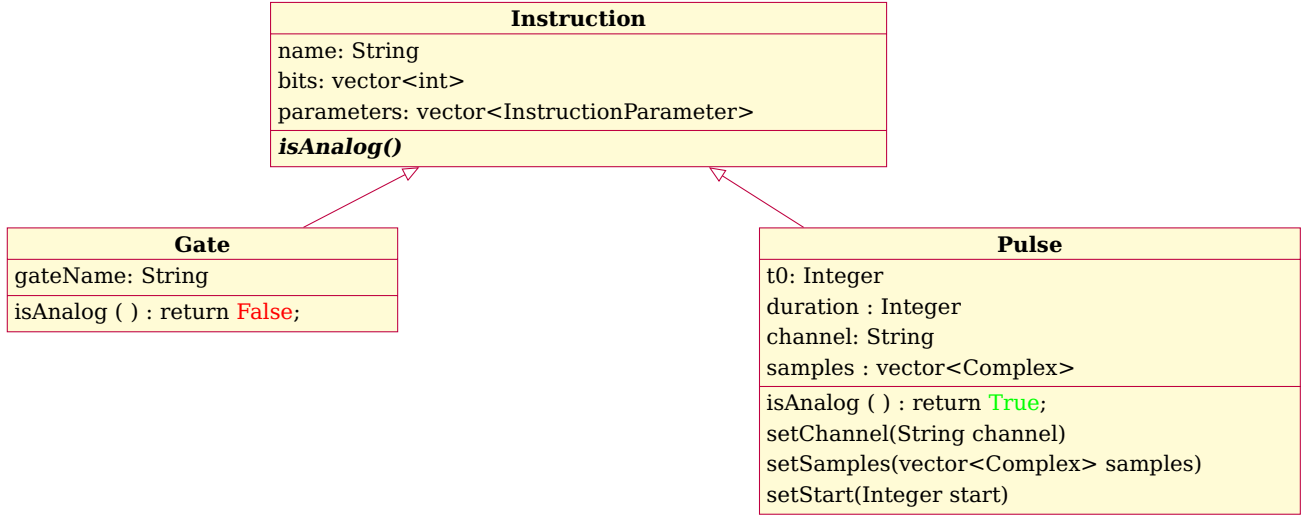
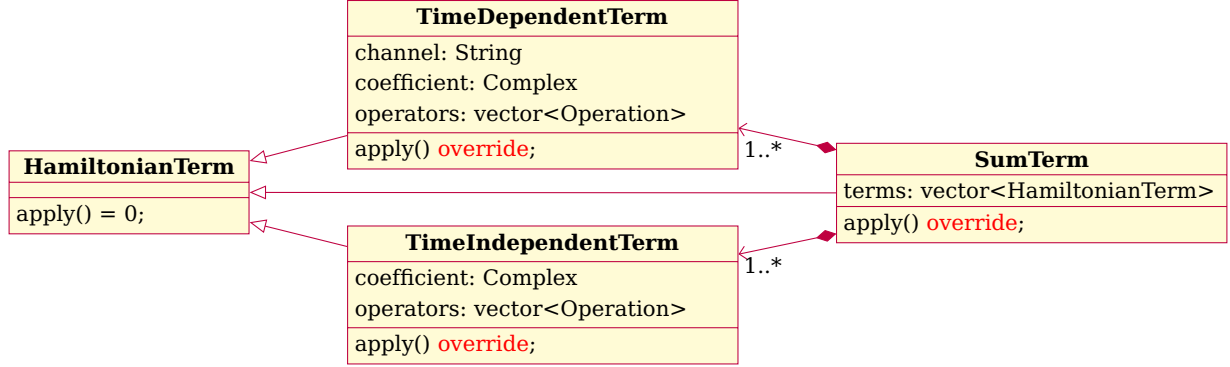
\begin{figure*}[h!] 
\begin{subfigure}{\textwidth}
  \begin{tikzpicture}
  \begin{class}[text width=8cm]{Instruction}{6 ,0}
  \attribute {name: String}
  \attribute {bits: vector<int>}
  \attribute {parameters:  vector<InstructionParameter>}
  \operation [0]{\textbf{isAnalog()}}
  \end{class}
  \begin{class}[text width =5 cm]{Gate}{1 , -3}
  \inherit{Instruction}
  \attribute {gateName: String}
  \operation {isAnalog ( ) : return \textcolor{red}{False}; }
  \end{class}
  \begin{class}[text width =7 cm]{Pulse}{12 , -3}
  \inherit {Instruction}
  \attribute {t0: Integer}
  \attribute {duration : Integer}
  \attribute {channel: String}
  \attribute {samples : vector<Complex>}
  \operation{isAnalog ( ) : return \textcolor{green}{True}; }
  \operation{setChannel(String channel)}
  \operation{setSamples(vector<Complex> samples)}
  \operation{setStart(Integer start)}
  \end{class}
  \end{tikzpicture}
  \caption{C++ class polymorphism is used to describe both quantum gate and pulse instructions.}
  \label{fig:instruction_polymorphism}
\end{subfigure}
\begin{subfigure}{\textwidth}
 \centering
  \begin{tikzpicture}
  \begin{class}[text width = 3cm]{HamiltonianTerm}{-2.5 ,0}
  \operation {apply() = 0;}
  \end{class}
  \begin{class}[text width = 5cm]{TimeIndependentTerm}{3, -1}
  \inherit{HamiltonianTerm}
  \attribute {coefficient: Complex}
  \attribute {operators: vector<Operation>}
  \operation {apply() \textcolor{red}{override};}
  \end{class}
  \begin{class}[text width = 5 cm]{TimeDependentTerm}{3, 2}
  \inherit{HamiltonianTerm}
  \attribute {channel: String}
  \attribute {coefficient: Complex}
  \attribute {operators: vector<Operation>}
  \operation {apply() \textcolor{red}{override};}
  \end{class}
  \begin{class}[text width = 5 cm]{SumTerm}{9.5, 0.1}
  \inherit {HamiltonianTerm}
  \attribute {terms: vector<HamiltonianTerm>}
  \operation {apply() \textcolor{red}{override};}
  \end{class}
  \composition{SumTerm}{}{1..*}{TimeDependentTerm}
  \composition{SumTerm}{}{1..*}{TimeIndependentTerm}
  \end{tikzpicture}
  \caption{C++ class structure represents different type of Hamiltonian terms.}
  \label{fig:hamiltonian_term_polymorphism}
\end{subfigure}
\caption{
\label{fig:xacc_uml_diagrams}%
Unified Modeling Language (UML) diagrams representing (a) IR \lstinline{Instruction} and \lstinline{Pulse} extension, and (b) Hamiltonian term polymorphism.}
\end{figure*}

In the XACC framework, the \lstinline{Instruction} and \\\lstinline{CompositeInstruction} intermediate representation (IR) interfaces are the building block of in-memory data structures capturing the semantics of a user-specified quantum program. 
%Thanks to C++ polymorphism, both individual Instruction objects and higher-level \texttt{CompositeInstruction} objects are manipulable via this uniform IR interface. 
In a usual setting, an \lstinline{Instruction} represents a quantum kernel instruction, i.e. a quantum gate. To handle analog-like pulse instructions, we define a \lstinline{Pulse} sub-class of \lstinline{Instruction} (Fig.~\ref{fig:xacc_uml_diagrams}~ \subref{fig:instruction_polymorphism}) 
that encapsulates extra metadata associated with analog control such as data samples, drive channel and pulse timing. We extend \lstinline{Instruction} with typical getter and setter methods for this metadata, which by default (for digital gate instructions) has no effect. We implement these methods on \lstinline{Pulse} to provide this metadata upon request. Moreover, we extend \lstinline{Instruction} with a \lstinline{isAnalog()} method to easily indicate if a set of  \lstinline{Instructions} are digital or analog. 
This allows us to construct flexible combinations of instructions (analog or digital) when building up a \lstinline{CompositeInstruction}, i.e. a quantum circuit/kernel.   

In pulse mode, analog IR objects can be constructed via direct calls to the IR factory methods (Fig.~\ref{fig:pulse_ir_c++} \& \ref{fig:pulse_ir_python}) or using XACC pulse library loader (Fig.~\ref{fig:pulse_ir_json_load}.)
The former offers maximal flexibility whereby arbitrary pulse instructions can be constructed from sample arrays. The latter provides a means for users to utilize any OpenPulse-conformed pulse library to build up their pulse program in XACC. Pre-defined and calibrated pulses are loaded into the instruction registry of the XACC IR provider for later use by referring to the pulse instruction name. In XACC, pulse composites that implement gates are prefixed with a '\lstinline{pulse::}' phrase. It is worth noting that, in pulse execution mode, XACC will automatically lower digital gate instructions to pulse composites as described in Section~\ref{sec:pulse-program-compilation}.
\begin{figure}[h!] 
\begin{subfigure}{.45\textwidth}
\lstset {language=C++}
\begin{lstlisting}
auto provider = xacc::getIRProvider("quantum");
auto compositeInst = provider->createComposite("test_pulse");
// Walsh-Gaussian pulse 
auto pulseWalshGaussian = 
        make_shared<Pulse>("WalshGaussian", "d0");
// Pulse reference: arxiv:1809.03452 (sec 7.1)
std::vector<std::vector<double>> samples = {
    {0.0, 0.0}, {0.013434, 0.0}, {0.058597, 0.0},  
    .....
};   
pulseWalshGaussian->setSamples(samples);
// Add the pulse instruction to the CompositeInstruction
compositeInst->addInstruction(pulseWalshGaussian);
\end{lstlisting}
\caption{Define a pulse via C++ IR manipulation.}
\label{fig:pulse_ir_c++}
\end{subfigure}
\begin{subfigure}{.45\textwidth}
\begin{python}
# Pulse data samples
pulseData = numpy.array([0+0j, 0.013434+0j, ...])
# Register the pulse instruction with XACC
xacc.addPulse('WalshGaussian', pulseData)   
provider = xacc.getIRProvider('quantum')
compositeInst = provider.createComposite('test_pulse')
pulseWalshGaussian = xacc.createPulse('WalshGaussian', 'd0')
pulseWalshGaussian.setBits([0])
compositeInst.addInstruction(pulseWalshGaussian)
\end{python}
\caption{Define a pulse via Python IR manipulation.}
\label{fig:pulse_ir_python}
\end{subfigure}
\begin{subfigure}{.45\textwidth}
\begin{python}
# Load backend pulse library JSON file 
with open('backends.json', 'r') as backendFile:
    jjson = backendFile.read()
# Use a pulse-capable accelerator
qpu = xacc.getAccelerator('ibm')
# Contribute pulse instructions 
# and pulse composites (cmd-def) 
# to the instruction registry
qpu.contributeInstructions(jjson);

provider = xacc.getIRProvider('quantum')
pulseCircuit = provider.createComposite('Bell')
# Retrieve pulse instructions from the IR provider
hadamard = provider.createInstruction('pulse::u2_0', 
    [0], [0.0, np.pi])
cnot = provider.createInstruction('pulse::cx_0_1', [0,1])
# Construct the pulse program
pulseCircuit.addInstructions([hadamard, cnot])
\end{python}
\caption{Pulse IR construction by loading pulse library JSON file. All pulses and pulse sequences (command definitions) are loaded to the IR service registry. These pulse IR's can then be retrieved from the IR provider.}
\label{fig:pulse_ir_json_load}
\end{subfigure}
\caption{\label{fig:pulse_ir} XACC pulse-level IR construction via (a) C++ API, (b) Python API, and (c) IR provider.}
\end{figure}

Besides being able to construct quantum kernels from pulse-level IR objects, the framework has extended its compilation infrastructure to handle automatic gate-to-pulse IR lowering. Hence, existing quantum circuits can be translated into pulse sequences and simulated dynamically within the framework itself. The existing XACC \lstinline{CompositeInstruction} API allows us to add, remove, or replace component Instruction nodes, which is crucial for the pulse-aware lowering of quantum kernels as we will describe in detail next.

\subsection{Pulse Program Compilation}\label{sec:pulse-program-compilation}
An important feature that we want to enable for XACC is the ability to transpile or compile gate or assembly level descriptions of quantum circuits into analog pulses that have been arranged to individual channels at specific time slots. This capability previously is only available on non-public hardware vendor backend software stacks. Since we want to implement an end-to-end pulse-capable stack for XACC that spans from front-end to back-end, including a pulse emulator backend that we will introduce in Section \ref{sec:quac}, we also implement a generic digital quantum gate to pulse compilation workflow.

Starting with an \lstinline{CompositeInstruction} tree structure, e.g. by compiling a quantum assembly program, XACC provides a flexible \lstinline{InstructionVisitor} interface to traverse this \lstinline{Instruction} tree and apply custom operations for each concrete \lstinline{Instruction} (gate) type. Following this visitor pattern, we have implemented a two-pass gate-to-pulse lowering and scheduling procedure as described in Algorithm~\ref{algo:pulse_visitor} \& \ref{algo:pulse_schedule}. 
\begin{algorithm} 
\SetAlgoLined
\KwData{Composite Instruction}
\KwResult{Composite Instruction contains only Pulse (analog) Instructions or Pulse Composites}
 \ForEach {Instruction of the Composite Instruction}{
  \If{is analog}{
    Skip\;
  }
  \Else{
    Look up pulse command definition\;
    \If{has cmd-def}{
        Replace gate with pulse composite\;
    }
    \Else{
        Decompose into universal gates\;
        Re-visit each component gate instruction\;
    }
  }
 }
 \caption{Gate-to-pulse compilation}
 \label{algo:pulse_visitor}
\end{algorithm}
\begin{algorithm}
\SetAlgoLined
\KwData{Pulse Composite Instruction}
\KwData{Start time}
\KwResult{Scheduled Pulse Composite Instruction}
 \ForEach {Instruction of the Composite Instruction}{
  \If{is NOT Composite}{
    Shift start time by parent composite start time\;
  }
  \Else{
    Calculate the current end time of scheduled pulses\;
    Re-call this procedure for the composite using the current end time as composite's start time\;
  }
 }
 \caption{Pulse Scheduling}
 \label{algo:pulse_schedule}
\end{algorithm}

In the first pass, all gates are transpiled into pulses or pulse sequences via \lstinline{Instruction} registry look-up. If a gate has a direct pulse command definition available, e.g. $X$ or $CNOT$ gates, it will be substituted by the associated pulse sequence. However, many quantum gates do not have direct pulse definitions, e.g. due to extended over-complete instruction sets. In that case, we rely on the universal instruction set that the registered pulse library implemented to decompose the gate into its universal gate equivalent and then lower those decomposed gates into pulses. After converting the original gate IR into universal gate set IR's, we revisit those IR nodes which should then have pulse command definitions required for the lowering. 

It is worth noting that pulse sequences are time-domain series all referencing time zero. These pulse sequences are then assembled as blocks in order to implement a specific quantum circuit. Therefore we must implement a compiler pass that schedules pulses globally. 
%To guarantee backend-agnostic, in the XACC framework, the whole pulse IR tree is the scheduled globally with respect to the circuit start time before submitting to the backend. 
This high-level pulse scheduling is described in Algorithm~\ref{algo:pulse_schedule}. To achieve this, we have extended XACC with a new \lstinline{Scheduler} service interface which we implement specifically for global pulse scheduling - the \lstinline{PulseScheduler}. 

This implementation basically block-shifts pulse-level \lstinline{CompositeInstructions}, i.e. those pulse sequences that implement quantum gates, by adjusting the start time of component pulses. This maintains the relative timing of pulses within a pulse sequence as well as the atomicity of quantum gates once they are converted into pulses. 

At the end of this procedure, we have a globally scheduled deck of pulse instructions, e.g. the one listed in Fig.~\ref{fig:quac_pulse_schedule}, dictating analog signals on channels driving the system Hamiltonian. This pulse program can hence be submitted to a pulse-capable hardware backend or to the QuaC simulator backend that we have provided as part of the XACC suite.
\begin{figure}[h!] 
  \begin{lstlisting}[language=json]
  [{"name":fc, "channel":d0, "time": 0, "duration": 0, "phase": -3.14159},
  {"name":X_PI_2_D0p, "channel":d0, "time": 0, "duration": 28},
  {"name":fc, "channel":d0, "time": 29, "duration": 0, "phase": -1.5708},
  {"name":X_PI_2_D0m, "channel":d0, "time": 29, "duration": 28},
  {"name":fc, "channel":d0, "time": 57, "duration": 0, "phase": -3.14159},
  {"name":X_PI_2_D0p, "channel":d0, "time": 57, "duration": 28},
  {"name":X_PI_2_D0m, "channel":d0, "time": 86, "duration": 28},
  {"name":fc, "channel":d0, "time": 114, "duration": 0, "phase": -3.14159},
  {"name":X_PI_2_D0p, "channel":d0, "time": 114, "duration": 28},
  {"name":fc, "channel":d0, "time": 143, "duration": 0, "phase": -1.5708},
  {"name":X_PI_2_D0m, "channel":d0, "time": 143, "duration": 28}]
  \end{lstlisting}
  \caption{Example of a pulse schedule. The \lstinline{time} field of each pulse is shifted by the pulse scheduler
  to construct a global pulse schedule after assembling individual pulse sequences.}
  \label{fig:quac_pulse_schedule}
\end{figure}

\subsection{QuaC Pulse Accelerator} \label{sec:quac}
The XACC framework defines an \lstinline{Accelerator} service 
that provides a unified interface for all framework services
that need access to a quantum computer, either
via a remote REST connection or local simulator. 
Therefore, we also implement our pulse-level simulator to adhere to this \lstinline{Accelerator}
interface in order to guarantee drop-in compatibility with existing services as illustrated in Fig.~\ref{fig:xacc_components}. In other words, existing projects 
written for the framework, e.g. for a quantum chemistry problem or a some other quantum algorithm, can immediately
utilize the pulse-capable simulator back-end without any modifications.
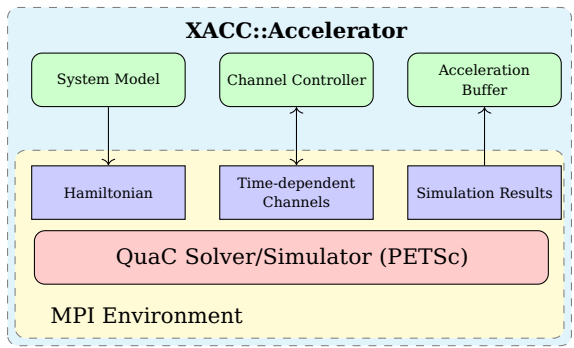
\begin{figure}[b!] 
\centering
%%% XACC QuaC block diagram %%%
% We need layers to draw the block diagram
\pgfdeclarelayer{background}
\pgfdeclarelayer{foreground}
\pgfsetlayers{background,main,foreground}

% Define a few styles and constants
\tikzstyle{acceleratorBlock} = [draw, fill=red!20, text width=20em, 
text centered, minimum height=2.5em, minimum width=24em, rounded corners]

\tikzstyle{component}=[draw, fill=blue!20, text width=6.5em, 
    text centered, minimum height=2.5em, font=\scriptsize	]

\tikzstyle{func}=[draw, fill=green!20, text width=6.5em, 
    text centered, minimum height=2.5em, font=\scriptsize, rounded corners]

\tikzstyle{ann} = [above, text width=5em]

\def\blockdist{2.0}
\def\edgedist{2.5}
\def\verticaldist{1.5}

\begin{tikzpicture}
    \node (QuaC) [acceleratorBlock] {QuaC Solver/Simulator (PETSc)};
    
    % Note the use of \path instead of \node at ... below. 
    \path (QuaC.140) + (-\blockdist, 0.5) node (ham) [component] {Hamiltonian};
    \path (QuaC.140) + (0.5, 0.5) node (chan) [component] {Time-dependent Channels};
    \path (QuaC.140) + (\blockdist + 1.0, 0.5) node (result) [component] {Simulation Results};
    \path (QuaC.south west) + (1.5,-0.4) node (MPI) {MPI Environment};

    \path (QuaC.140) + (-\blockdist, 0.5 + \verticaldist) node (model) [func] {System Model};
    \path (QuaC.140) + (0.5, 0.5 + \verticaldist) node (controller) [func] {Channel Controller};
    \path (QuaC.140) + (\blockdist + 1.0, 0.5 + \verticaldist) node (buffer) [func] {Acceleration Buffer};

    \path [draw, ->] (model) -- node [right] {} (ham);
    \path [draw, <->] (controller) -- node [right] {} (chan);
    \path [draw, <-] (buffer) -- node [right] {} (result);

    \begin{pgfonlayer}{background}
        % Compute a few helper coordinates
        \path (QuaC.west |- MPI.south) + (-0.35, -0.2) node (a) {};
        \path (buffer.north -| QuaC.east) + (0.3, 0.6) node (b) {};
        \path[fill=cyan!10,rounded corners, draw=black!50, dashed]
            (a) rectangle (b);

        \path (QuaC.west |- MPI.south) + (-0.25, -0.1) node (a) {};
        \path (result.north -| QuaC.east) + (0.2, 0.2) node (b) {};
        \path[fill=yellow!20,rounded corners, draw=black!50, dashed]
            (a) rectangle (b);
    \end{pgfonlayer}
    
    \path (controller.north) + (0.0, 0.3) node (xacc) {\textbf{XACC::Accelerator}};
\end{tikzpicture}
%%% XACC QuaC block diagram %%%
\caption{Block diagram of the XACC pulse simulator back-end based on QuaC.}
\label{fig:quac_sim_arch}
\end{figure}

The overall architecture of the simulator is illustrated in Fig.~\ref{fig:quac_sim_arch}. Unlike other circuit-based back-ends,
a pulse-level back-end must be aware of the quantum system dynamics, i.e. the Hamiltonian, and also be able to generate continuous time-dependent signals to drive the Hamiltonian according to the input gate or pulse instructions. In the following sub-sections, we will describe these pulse-specific components in greater detail.

\subsubsection{Hamiltonian Parser} \label{hamiltonian-parser}
As previously mentioned, we adopt the OpenPulse specification, which
standardizes various data formats for pulse-level control. With respect to the quantum hardware system dynamics,
as summarized in Section~\ref{hamiltonian}, the Hamiltonian is expressed as a list of terms,
each of which can be time-independent, time-dependent (driven by a channel) or an indexed sum of 
similar terms. 

C++ polymorphism provides an effective way to capture the Hamiltonian abstraction as depicted in Fig.~\ref{fig:hamiltonian_term_polymorphism}
whereby different derived sub-classes of a common abstract class handle concrete types of Hamiltonian terms.
Custom actions on the system dynamics which are specific to each type of term are encapsulated in an \lstinline{apply()} method
whose implementation is provided by each child class. Based on the textual signature of each input term, the parser will construct an
appropriate object capturing all information specific to the type of the term. The \lstinline{apply()} action will then reconstruct
the dynamics within the simulator, i.e. the Hamiltonian matrix.
\begin{algorithm}[!b] 
\SetAlgoLined
\KwData{Scheduled Pulse Program}
\KwData{Current Time}
\KwResult{Update Channel Active Pulse Sequence}
\KwResult{Update Channel Accumulated FC Phase}
 \ForEach {Active Pulse on Channel}{
  \If{still active}{
    Skip\;
  }
  \Else{
    Remove current Pulse on Channel\;
    Find next Pulse scheduled\;
    Update current Pulse on Channel\; 
  }
 }
 \ForEach {Channel}{
    \ForEach {Frame Change instructions on channel}{
        \If{Activated @ current time} {
            Update accumulated Phase on Channel\;
            Remove processed Frame Change instruction\; 
        }
    }
 }
 \caption{Channel Controller Tick algorithm: updating its internal state during time-stepping.}
 \label{algo:channel-controller}
\end{algorithm}

\subsubsection{Pulse Controller} \label{pulse-controller}
As depicted in Fig.~\ref{fig:quac_sim_arch}, instantaneous values of all driving time-dependent terms are
provided by a centralized \lstinline{ChannelController} whose task is to process the scheduled pulse program
to figure out the driving signal values on all channels. This involves (1) tracking which pulse is active on a given 
channel at a given time and how far along its data sample list we are, (2) accounting for all frame changes (local oscillator phase manipulation) 
that have requested on any given channels, and (3) performing LO-mixing to compute the absolute time signals. The way in which
our pulse controller handles these tasks is summarized in Algorithm~\ref{algo:channel-controller}.   

\subsubsection{QuaC Simulator} \label{Simulator}
The numerical back-end of our implementation, so-called Quantum in C - \lstinline{QuaC} \cite{quac-github}, as shown in Fig.~\ref{fig:quac_sim_arch} 
is a quantum dynamical solver based on the Portable, Extensible Toolkit for Scientific Computation (PETSc) library~\cite{abhyankar2018petsc, balay2019petsc}.
Built upon PETSc's scalable (parallel) data-structures and runtimes, the \lstinline{QuaC} back-end in XACC supports multi-node
parallelism via the industry-standard Message Passing Interface (MPI), which promises extended scalability across classical computing infrastructure.

As a standalone package, QuaC provides a flexible open quantum systems simulator capable of solving generic Lindblad master equations~(\ref{eq:master_equation}) as well as applying gate-type operators. 
\begin{equation}
\begin{split}
\frac{d\rho}{dt} &= -i[H, \rho] + \sum_k \gamma_k (A_k \rho A_k^\dagger - \frac{1}{2}A_k^\dagger A_k \rho  - \frac{1}{2}\rho A_k^\dagger A_k) \\
 &= \mathcal{L}(\rho)
\end{split}
\label{eq:master_equation}
\end{equation}

The vectorized density matrix ($\rho$) and the superoperator ($\mathcal{L}$) representing the right-hand side of the master equation are modeled by the abstract PETSc vector (\lstinline{Vec}) and matrix (\lstinline{Mat}) objects whose implementation details, such as memory allocation and storage across multiple MPI processes, are handled implicitly by PETSc for maximal portability. Specifically, the superoperator matrix is stored in the compressed sparse row (CSR) format across multiple (parallel) processes using the PETSc-builtin MPI AIJ type.

Qubit sub-system operators, e.g. Pauli operators, which can be used in the OpenPulse Hamiltonian, are implicitly modeled internally as functions that retrieve non-zero matrix elements and indices to take advantage of their sparsity. These functions are used to project operators' local elements into the overall RHS superoperator matrix when we process Hamiltonian terms, i.e. executing the \lstinline{apply()} method (Fig.~\ref{fig:hamiltonian_term_polymorphism}) on each term. All channel-driven terms in the Hamiltonian are modeled as a time-dependent function to compute the Jacobian which is a PETSc callback at each time step.

It is worth noting that PETSc offers a wide variety of time integrators, such as forward and backward Euler methods, symplectic methods, backward differentiation formula (BDF) method, etc. By default, QuaC uses \lstinline{TSRK3BS} integrator type (third-order Runge-Kutta scheme of Bogacki-Shampine with 2nd order embedded method) when solving the master equation. 

As depicted in Fig.~\ref{fig:quac_sim_arch}, all these low-level numerical computation details are encapsulated within the \lstinline{xacc::Accelerator} interface thanks to the high-level system model, pulse channel controller, and accelerator buffer data structures. The first and second objects translate the input Hamiltonian, backend configurations (e.g. LO frequencies), and the pulse program into QuaC and PETSc data structures and functions required for solving the pulse-driven system dynamics. The third object (a core abstraction from XACC modeling a register of qubits and exposing pertinent backend-execution results) captures all necessary outputs from the simulation such as the probability distribution which can be used to generate a measurement shot count distribution.

\subsubsection{Pulse Generator Utility} \label{sec:pulse_util}
In an effort to improve the usability and ease of integration, we also put forward a set of utility functions that generate pulse data series for common pulse shapes, such as square or Gaussian pulses. Given the analytical parameters of the envelope function, e.g. the width (variance) parameter of a Gaussian function, the sampling time ($dt$), and the pulse duration, these utilities will generate the corresponding array of data samples. In the most generic form, users can also supply a string-based functional expression which will then be parsed and evaluated to compute the time-domain samples.

These pulse generation utilities are also made available to the Python language binding for Pythonic programming. Python has a rich collection of signal processing libraries, e.g. \lstinline{scipy.signal}, which can be used independently or in conjunction with these XACC-provided facilities to construct input pulses. 

\section{Demonstration} \label{Demonstration}
In this section, we seek to demonstrate the usage of XACC pulse-level application programming interfaces to accomplish tasks ranging from simple physics experiments to complex, cross-service applications involving integration of new and existing XACC components.

\subsection{Arbitrary pulse definition}
\label{sec:Example1}
As a first example, we describe how arbitrary pulse waveforms can be described via direct construction and manipulation of pulse-level IR instances (see Fig.~\ref{fig:quac_pulse_ir_snippet}).
In particular, the \lstinline{setSamples} method of the \lstinline{Pulse} class (see also Fig.~\ref{fig:xacc_uml_diagrams}\subref{fig:instruction_polymorphism}) 
can be used to attach a list of sample data points to a pulse IR. The \lstinline{xacc::getAccelerator} API function enables one to retrieve an \lstinline{Accelerator}
back-end by name from the framework service registry. The \lstinline{CompositeInstruction}, i.e. quantum program,
which contains pulse-level instructions can be simulated/executed on the back-end by invoking the \lstinline{execute} method and passing references to the 
program and the \lstinline{AccelerationBuffer} (see Fig.~\ref{fig:quac_sim_arch}) to which the results are persisted.

\begin{figure}[t!] 
\lstset {language=C++}
\begin{lstlisting}
  auto quaC = xacc::getAccelerator("QuaC");    
  // Create a simple pulse program using IR
  auto provider = xacc::getIRProvider("quantum");
  auto compositeInst = provider->createComposite("test_pulse");
  // Walsh-Gaussian pulse 
  auto pulseWalshGaussian = 
            make_shared<Pulse>("WalshGaussian", "d0");
  // Pulse reference: https://arxiv.org/abs/1809.03452 (sec 7.1)
  std::vector<std::vector<double>> samples = {
        {0.0, 0.0}, {0.013434, 0.0}, {0.058597, 0.0},  
        .....
  };   
  pulseWalshGaussian->setSamples(samples);

  // Add the pulse and execute
  compositeInst->addInstruction(pulseWalshGaussian);

  auto qubitReg = xacc::qalloc(1);    
  // Run simulation
  quaC->execute(qubitReg, compositeInst);
\end{lstlisting}
\caption{Define a pulse via IR (Intermediate Representation) manipulation.}
\label{fig:quac_pulse_ir_snippet}
\end{figure}

\subsection{Finding resonance frequencies}

In this example, we demonstrate the way the system Hamiltonian is specified (Fig.~\ref{fig:quac_hamiltonian_snippet}) and
how to configure various simulation parameters to experiment on- and off-resonance driving of qubits.
\begin{figure}[!b]
\begin{lstlisting}[language=json]
{
    "description": "Qubits are modelled as a two level system. System of 2 qubits.",
    "h_str": ["_SUM[i,0,1,wq{i}/2*Z{i}]", "_SUM[i,0,1,omegad{i}*X{i}||D{i}]", "jq0q1*Sp0*Sm1", "jq0q1*Sm0*Sp1"],
    "osc": {},
    "qub": {
        "0": 2,
        "1": 2
    },
    "vars": {
        "omegad0": 1.303125,
        "omegad1": 0.97, 
        "wq0": 30.91270129264568,
        "wq1": 30.36010168900955,
        "jq0q1": 0.04
    }
}
\end{lstlisting}
\caption{Hamiltonian JSON snippet for resonance frequency experiment.}
\label{fig:quac_hamiltonian_snippet}
\end{figure}
\begin{figure}
  \lstset {language=C++}
  \begin{lstlisting}
auto xasmCompiler = xacc::getCompiler("xasm");    
  
// X gate on the first qubit (sweeping LO freq)
auto ir1 = xasmCompiler->compile(R"(__qpu__void testQ1(qbit q)
{
  X(q[0]);
  Measure(q[0]);
})");    
auto program1 = ir1->getComposite("testQ1");

// X gate on the second qubit (sweeping LO freq)
auto ir2 = xasmCompiler->compile(R"(__qpu__ void testQ2(qbit q)
{
  X(q[1]);
  Measure(q[1]);
})");    
auto program2 = ir2->getComposite("testQ2");

// Sweeping Q1 freq
auto freq_q1 = xacc::linspace(30.7, 31.0, 50);
for (const auto& freq : freq_q1) {
   // Set the sweeping LO freq on D0 (first channel)
   std::vector<double> loFreqs {
      freq,
      30.36010168900955
   };
   // Set the channel configurations
   // to reflect the LO frequency
   BackendChannelConfigs channelConfigs;
   channelConfigs.loFregs_dChannels = loFreqs;
   systemModel->setChannelConfigs(channelConfigs);
   const int NB_SHOTS = 10000;
   auto quaC = xacc::getAccelerator("QuaC",
     std::make_pair("system-model", systemModel), 
     std::make_pair("shots", NB_SHOTS) });

   auto qubitReg = xacc::qalloc(2);
   // Run the Pulse simulation with the Hamiltonian provided
   quaC->execute(qubitReg, program1);
   // Store the probability of (measurement = 1)
   resultArray.emplace_back(
    qubitReg->computeMeasurementProbability("1"));
}
// Sweeping Q2 freq
auto freq_q2 = xacc::linspace(30.15, 30.45, 50);
...
// Similar to sweeping Q1, 
// now, we sweep the LO freq. of D1 (second channel)
  \end{lstlisting}
  \caption{XACC code snippet for sweeping qubit's LO frequency. In this example, the input is an assembly function (XASM dialect) which will be eventually lowered down to
  pulses whose time-domain values will be mixed with a local-oscillator signal to 
  drive the system Hamiltonian.}
  \label{fig:quac_freq_sweep_snippet}
\end{figure}  

\begin{figure}[h!] 
\begin{tikzpicture}[scale=1.0][domain=0:8]
\begin{axis}[
      xmin = 30.15, xmax = 31.06,
      xlabel = { Frequency ($2\pi \times GHz$) }, 
      ylabel = { Probability $|1\rangle$ }, 
      title = {Qubit 1 ($\omega_1 = 30.36$); Qubit 0 ($\omega_0 = 30.9127$)}]
\addplot[only marks, scatter]
      table[x=x, y=y] {
x       y
30.7	0.0019
30.7061	0.0034
30.7122	0.0034
30.7184	0.0038
30.7245	0.0031
30.7306	0.0033
30.7367	0.0035
30.7429	0.0047
30.749	0.0045
30.7551	0.0047
30.7612	0.0043
30.7673	0.0062
30.7735	0.0083
30.7796	0.0073
30.7857	0.0104
30.7918	0.0103
30.798	0.0122
30.8041	0.0117
30.8102	0.0129
30.8163	0.0133
30.8224	0.0205
30.8286	0.0308
30.8347	0.0436
30.8408	0.0784
30.8469	0.1163
30.8531	0.1732
30.8592	0.2502
30.8653	0.3408
30.8714	0.4492
30.8776	0.5544
30.8837	0.662
30.8898	0.7677
30.8959	0.8547
30.902	0.9207
30.9082	0.959
30.9143	0.9689
30.9204	0.9458
30.9265	0.8945
30.9327	0.813
30.9388	0.7188
30.9449	0.6162
30.951	0.5049
30.9571	0.3813
30.9633	0.2993
30.9694	0.2176
30.9755	0.1501
30.9816	0.0958
30.9878	0.0645
30.9939	0.0382
31	    0.0242
};
\addplot[only marks, scatter]
            table[x=x, y=y] {
      x       y
      30.15	0.0026
      30.1561	0.0022
      30.1622	0.002
      30.1684	0.0029
      30.1745	0.0033
      30.1806	0.0027
      30.1867	0.0028
      30.1929	0.0028
      30.199	0.0037
      30.2051	0.0038
      30.2112	0.0028
      30.2173	0.0057
      30.2235	0.0061
      30.2296	0.0063
      30.2357	0.0069
      30.2418	0.0077
      30.248	0.009
      30.2541	0.008
      30.2602	0.0117
      30.2663	0.0132
      30.2724	0.0198
      30.2786	0.0261
      30.2847	0.0444
      30.2908	0.074
      30.2969	0.1115
      30.3031	0.1593
      30.3092	0.2371
      30.3153	0.3203
      30.3214	0.4202
      30.3276	0.5275
      30.3337	0.6304
      30.3398	0.734
      30.3459	0.8244
      30.352	0.9038
      30.3582	0.9491
      30.3643	0.9678
      30.3704	0.9522
      30.3765	0.9094
      30.3827	0.8329
      30.3888	0.7439
      30.3949	0.625
      30.401	0.5322
      30.4071	0.4131
      30.4133	0.311
      30.4194	0.2218
      30.4255	0.1409
      30.4316	0.0972
      30.4378	0.0526
      30.4439	0.0336
      30.45	0.0193    
      };
\end{axis}
\end{tikzpicture}
\caption{Result of frequency-sweep experiment for qubit 0 (right peak) and qubit 1 (left peak).}
\label{fig:quac_freq_sweep_results}
\end{figure}
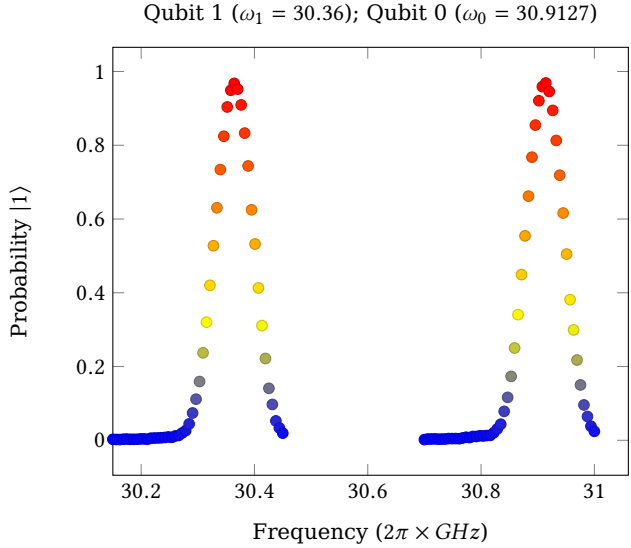

Specifically, we will apply an X gate (in \lstinline{XASM} dialect, Fig.~\ref{fig:quac_freq_sweep_snippet}), 
which will be translated to a corresponding pulse, on each qubit. Since the X-gate pulse will actually be mixed with the LO signal before being applied to the qubit,
we can vary the LO frequency in order to find the qubit resonance frequency.
In the below simulation, we sweep the frequency of the local oscillators of drive channel 0 (line 20, Fig.~\ref{fig:quac_freq_sweep_snippet}),  
and drive channel 1 (line 45, Fig.~\ref{fig:quac_freq_sweep_snippet}), where the X-gate pulses are applied and collect the probability that
the corresponding qubit is measured in the $|1\rangle$ state (line 42, Fig.~\ref{fig:quac_freq_sweep_snippet}.) The results are shown in Fig.~\ref{fig:quac_freq_sweep_results} which, as expected,
verify the resonance characteristics of the back-end Hamiltonian.

This example also demonstrates the end-to-end gate-to-pulse transformation capabilities of the XACC framework. 
In particular, rather than needing users to manually specify the pulses, which they can certainly do as in \ref{sec:Example1}, 
digital gates are automatically lowered into pulses and dynamically simulated once the \lstinline{QuaC} pulse back-end is requested. 
In other words, other quantum assembly language based circuits can also be simulated at the pulse level targeting a
specific hardware model via the \lstinline{QuaC} simulator back-end.

\begin{figure}[!b] 
\begin{lstlisting}[language=json]
{
    "description": "One-qutrit Hamiltonian.",
    "h_latex": "",
    "h_str": ["(w - 0.5*alpha)*O0", "0.5*alpha*O0*O0", "O*(SM0 + SP0)||D0"],
    "osc": {},
    "qub": {
        "0": 3
    },
    "vars": {
        "w": 31.63772297724,
        "alpha": -1.47969,
        "O": 0.0314
    }
\end{lstlisting}
\caption{Qutrit Hamiltonian JSON snippet. We explicitly specify the dimension of qubit subsystem 0 as 3 to model it as a qutrit.}
\label{fig:qutrit_hamiltonian}
\end{figure}
\subsection{Qutrit simulation}
The XACC pulse-level infrastructure that we have implemented is hardware-agnostic in that it supports not only two-level systems (qubits) but also arbitrary-dimension systems such as qutrits and qudits. In addition, the ability to model higher-dimensional systems allows us to accurately simulate qubit systems in their extended Hilbert space beyond the computational qubit subspace. 

\begin{figure}[!t]
\centering  
\includegraphics[width=0.5\textwidth]{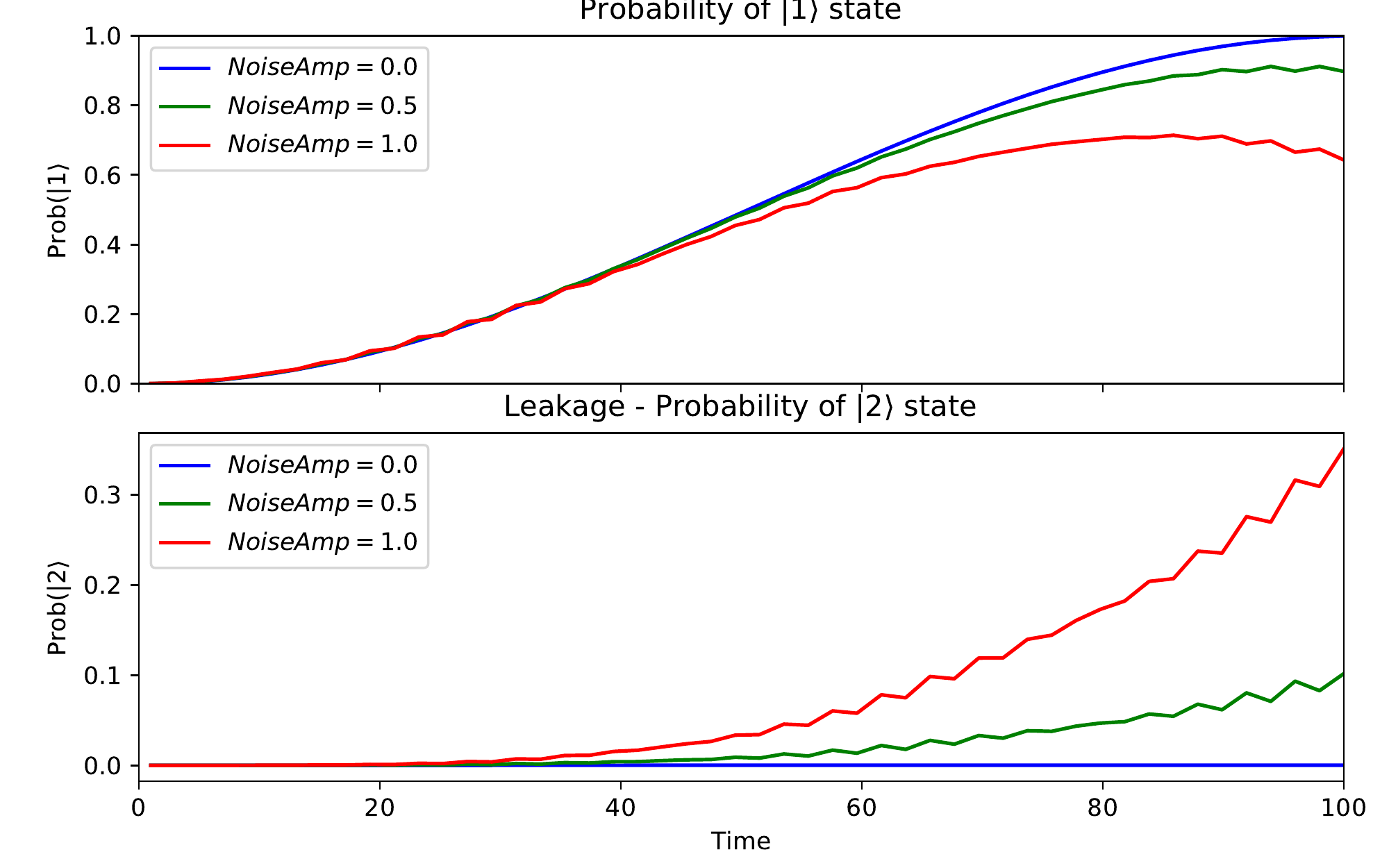}
\caption{Simulation result of a transmon qubit (modeled as a three-level system) driven by a $pi$-pulse plus noise (top) Probability of the |1> state vs. time (bottom) Leakage: probability of the |2> state vs. time. As we can see, the amount of leakage increase with the noise amplitude (at anharmonicity frequency.)}
\label{fig:qutrit_sim_result}
\end{figure}

For example, we can model superconducting transmon qubits as three-level systems (eq.~(\ref{eq:transmon_H})) in order to compute the amount of leakage into the undesirable (third) basis state.
\begin{equation}
    H = \Omega(t)(\hat{a} + \hat{a}^\dagger) + \omega_{01}\hat{n} + \frac{\alpha}{2}(\hat{n} - 1)\hat{n}
    \label{eq:transmon_H}
\end{equation}
Theoretically, the unwanted (leakage) transition ($|1\rangle \rightarrow |2\rangle$) is separated from the principle transition ($|0\rangle \rightarrow |1\rangle$) by the anharmonicity ($\alpha$), i.e. $\omega_{12} = \omega_{01} + \alpha$. Consequently, noise or Fourier-components of the driving signal at that anharmonicity frequency will result in leakage. We demonstrate this leakage by driving a transmon qubit which is described by the OpenPulse Hamiltonian in Fig.~\ref{fig:qutrit_hamiltonian} with a square $\pi$-pulse plus an additional fictitious noise term of the form, $Amplitude\times cos(\alpha t)$. By varying this $Amplitude$ parameter, we can examine the leakage while driving the system from $|0\rangle$ to $|1\rangle$. Fig.~\ref{fig:qutrit_sim_result} directly demonstrates this effect, and we see that increasing noise levels lead to expected leakage into the $|2\rangle$ state.

\subsection{Pulse optimization}
XACC is expanding on the QCOR Optimizer concept \cite{mintz2019qcor} for pulse-level, optimal quantum control. We leave a detailed description of this capability for a future manuscript, but want to demonstrate its utility here for pulse-level programming and compilation. Fig. \ref{fig:xacc_goat_optim} demonstrates how one may use standard optimal control techniques (e.g. GOAT, GRAPE, CRAB, etc.) to create a pulse-level program that affects targeted unitary evolution under a certain set of input parameters. This snippet shows how one might leverage this to optimize an arbitrary truncated Fourier series in order to implement an $X$ gate. 
\begin{figure}[h!] 
  \lstset {language=C++}
  \begin{lstlisting}
xacc::HeterogeneousMap configs {
    std::make_pair("method", "GOAT"),
    std::make_pair("optimizer", "ml-pack"),
    std::make_pair("dimension", dimension),
    // Target unitary is an X gate on qubit 0
    std::make_pair("target-U", "X0"),
    // Control parameter (used in the control function)
    std::make_pair("control-params", { 
    "a0", "a1", "a2", "a3", "a4", "a5", "a6", "a7", "a8", 
    "b1", "b2", "b3", "b4", "b5", "b6", "b7", "b8" }),
    // Fourier series (first 8 terms only)
    std::make_pair("control-funcs", { 
    "a0 + 
    a1*cos(1*0.1*t) + b1*sin(1*0.1*t) + 
    a2*cos(2*0.1*t) + b2*sin(2*0.1*t) + 
    a3*cos(3*0.1*t) + b3*sin(3*0.1*t) + 
    a4*cos(4*0.1*t) + b4*sin(4*0.1*t) + 
    a5*cos(5*0.1*t) + b5*sin(5*0.1*t) + 
    a6*cos(6*0.1*t) + b6*sin(6*0.1*t) + 
    a7*cos(7*0.1*t) + b7*sin(7*0.1*t) + 
    a8*cos(8*0.1*t) + b8*sin(8*0.1*t)" }),
    // The list of Hamiltonian terms 
    // that are modulated by the control functions
    std::make_pair("control-H", { "X0" }),
    // Initial params
    std::make_pair("initial-parameters", initParams),
    std::make_pair("max-time", tMax)
};
auto optimizer = xacc::getOptimizer("quantum-control", configs);
auto result = optimizer->optimize();
auto optimal_a_b_vec = result.second;
\end{lstlisting}
\caption{Using XACC pulse optimization service (Gradient optimization of analytic controls - GOAT) to optimize a waveform given by a Fourier series (lines 13-21) to implement a target unitary (line 6) which can be given as a text input as in this example or as a matrix.}
\label{fig:xacc_goat_optim}
\end{figure}
This is useful
when, e.g., we want to constrain the bandwidth of the drive signal.

\begin{figure}[h!]
\centering  
\includegraphics[width=0.5\textwidth]{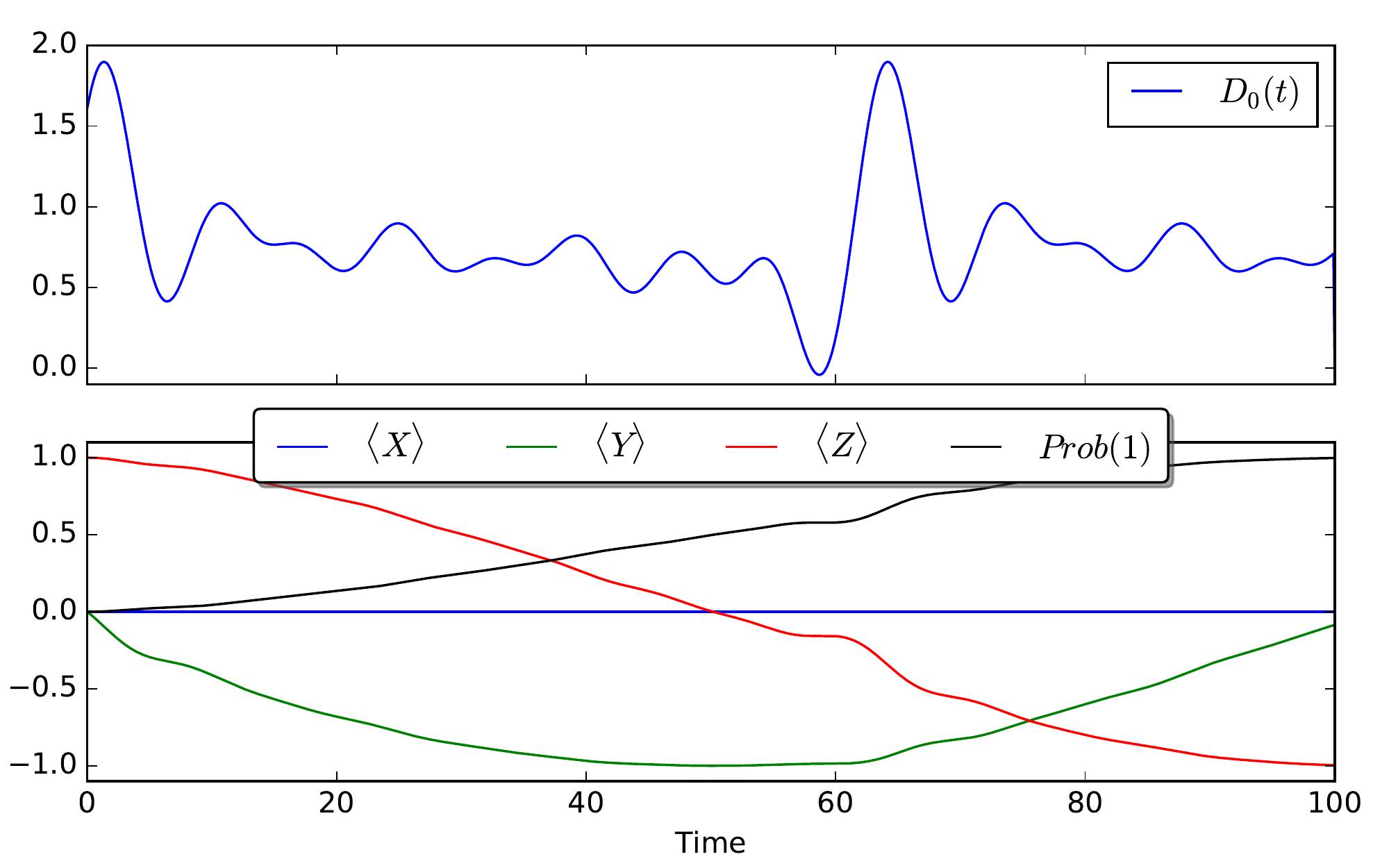}
\caption{Simulation result of the Fourier-series pulse: (top) drive signal envelop and (bottom) expectation values of Pauli observables and number operator.}
\label{fig:quac_pulse_sim_result}
\end{figure}

The optimal parameters that were found by the XACC pulse optimization plugin can then be verified on the QuaC backend as shown in the Python script in Fig.~\ref{fig:quac_pulse_analytical}.   
\begin{figure}[h!] 
\begin{python}
fourierSeries = '''
    0.726924 + 
    0.065903*cos(1*0.1*t) + 0.128627*sin(1*0.1*t) +
    0.079360*cos(2*0.1*t) + 0.111686*sin(2*0.1*t) + 
    0.096717*cos(3*0.1*t) + 0.096822*sin(3*0.1*t) + 
    0.106937*cos(4*0.1*t) + 0.092216*sin(4*0.1*t) + 
    0.215306*cos(5*0.1*t) + 0.118562*sin(5*0.1*t) +
    0.117682*cos(6*0.1*t) + 0.126134*sin(6*0.1*t) + 
    0.100447*cos(7*0.1*t) + 0.120409*sin(7*0.1*t) + 
    0.103292*cos(8*0.1*t) + 0.108712*sin(8*0.1*t)'''

channelConfigs.addOrReplacePulse('fourier', 
    xacc.PulseFunc(fourierSeries, nSamples, channelConfigs.dt))
model.setChannelConfigs(channelConfigs)

qubitReg = xacc.qalloc(1)

provider = xacc.getIRProvider('quantum')
composite = provider.createComposite('test_pulse')
pulse = xacc.createPulse('fourier', 'd0')
composite.addInstruction(pulse)

qpu.execute(qubitReg, composite)
\end{python}
  \caption{Example (Python) of pulse construction by analytical expressions.}
  \label{fig:quac_pulse_analytical}
\end{figure}
The drive signal (analytical form), line 1-10, which is the result from the pulse optimizer, can be converted to a pulse sequence using
the pulse generation utility (line 13) that we have described in~\ref{sec:pulse_util} . It can be then verified by the simulator back-end
to confirm that the optimized pulse indeed implemented an $X$ gate. Fig.~\ref{fig:quac_pulse_sim_result} shows the time-domain data
resulting from executing the pulse instruction in Fig.~\ref{fig:quac_pulse_analytical}.  

As we can see, the probability of $|1\rangle$ state and the expectation value of the $Z$ operator
can confirm that this pulse indeed implement an $X$ gate. 

Within the XACC framework, there are many modular and well-defined services that users can leverage for pulse-level experiments.
For instance, the framework provides various standardized optimization and machine-learning services which can handle 
pulse-level optimal control problems. In a similar vein, by adhering to common framework interfaces,
pulse-level services, especially the pulse-capable back-end, that we implement in this work, are inter-operable with all other framework components that are listed in Fig.~\ref{fig:xacc_components}.

\subsection{MPI Demonstration}
Finally, as explained in Section \ref{Simulator}, the QuaC Accelerator simulation backend is built upon the standard MPI runtime environment (through PETSc), and can therefore be easily parallelized to suit the underlying classical compute architecture. Here, we seek to demonstrate that capability by checking the simulation runtime under various MPI parallelization configurations, specifically the number of MPI processes.

\begin{figure}[h!]
\centering  
\includegraphics[clip, trim=2cm 4cm 2cm 4cm, width=0.5\textwidth]{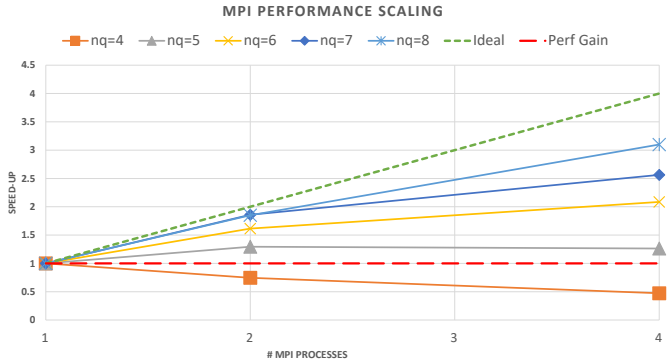}
\caption{Performance scaling vs. number of MPI processes. The bigger the system size (in terms of the number of qubits, \lstinline{nq}) is, the closer to ideal (linear) performance scaling we can achieve. When simulating small systems, e.g. less than 5 qubits, it's best to use the single processing mode. Hardware descriptions: desktop workstation, Core i7-8700, 6 (12) cores (threads).}
\label{fig:mpi_result}
\end{figure}

The results of these scaling experiments are shown in Fig.~\ref{fig:mpi_result} where we have normalized the runtime of the single-process configuration as one. In this test, we vary the system size in terms of the number of qubits and perform a pulse simulation whereby all qubits are driven by an arbitrary pulse. 

There are two scaling regimes of interest, (1) the perfect, linear scaling denoted by the green dash line in Fig.~\ref{fig:mpi_result} and (2) the zero-scaling horizontal line at 1.0 (the red dash line in Fig.~\ref{fig:mpi_result}.) Anything below the horizontal line (2) is not suitable for MPI multi-processing, i.e. the inter-process communication and synchronization overheads exceed any parallel-processing gain. 

In this regard, we can see that systems of small sizes (e.g. 4-5 qubits) are best to be simulated using a single MPI process. On the other hand, large systems can take advantage of MPI parallelism with a considerable amount of speed-up. The larger the system size is, the closer we can get to the ideal scaling regime.  

\section{Discussion and Future Work}\label{Discussion}
% \subsection{Summary} 
We have extended the XACC framework to implement a unified end-to-end programming model 
consisting of multiple layers of abstraction from high-level constructs of algorithms and observables
via gate-level representations to pulse-level decomposition. Thanks to this extension, entry-level users can experiment
with dynamical simulation or pulse-level execution of their quantum applications without needing to change any of their code.
On the other hand, by defining the common pulse-related API's for the framework, professional users can access
those API's to provide custom, high-performance implementations, e.g. a custom pulse library and gate to pulse mapping,
a specialized pulse scheduler, or a performant pulse-level simulator, and contribute those implementations as services to the framework.
This work provides a much needed capability in the nascent quantum programming, compilation, and execution research landscape. We anticipate 
this work to enable new efforts with regards to quantum optimal-control, error mitigation and noise-suppression, and the development of novel 
multi-qubit pulse sequences for physical system simulation.

% \subsection{Future Work} 
We note that pulse-level programming is an emerging field which we believe will certainly experience rapid iteration and advancement.
Hence, our immediate future goals are to keep the framework relevant to latest industry standards and trends.
In particular, we have plans to
\begin{enumerate}
\item Adopt other pulse-level programming models as they emerge, e.g., Rigetti's \lstinline{Quilt}, in a continuous integration fashion.
\item Enable translation (where applicable) between pulse-level program representations across physical hardware implementations.
\item Expand and improve our pulse-related service capabilities such as applying quantum optimal control to pulse synthesis and optimizing quantum
programs at pulse level (e.g. pulse scheduling). 
\item Develop standardized pulse-level emulator environments which include high-quality, fully-calibrated quantum hardware system models and pulse libraries
for learning and testing purposes. 
\end{enumerate}

\section*{Acknowledgements}
\label{}
This work has been supported by the US Department of Energy (DOE) Office of Science Advanced Scientific Computing Research (ASCR) Quantum Computing Application Teams (QCAT), Quantum Algorithms Team (QAT), and Accelerated Research in Quantum Computing (ARQC). ORNL is managed by UT-Battelle, LLC, for the US Department of Energy under contract no. DE-AC05-00OR22725. The US government retains and the publisher, by accepting the article for publication, acknowledges that the US government retains a nonexclusive, paid-up, irrevocable, worldwide license to publish or reproduce the published form of this manuscript, or allow others to do so, for US government purposes. DOE will provide public access to these results of federally sponsored research in accordance with the DOE Public Access Plan.

\bibliographystyle{unsrt}
\bibliography{main}

\end{document}